\input harvmac
\sequentialequations
\def\hphi{\hat{\phi}}
\def\a{\alpha}
\def\b{\beta}
\def\g{\gamma}
\def\d{\delta}
\def\p{\partial}
\lref\mtheory{P. Townsend, Phys. Lett. {\bf 350B} (1995) 184, hep-th/9501068.}
\lref\mtheoryt{E. Witten, Nucl. Phys. {\bf B443} (1995) 85, hep-th/9503124.}
\lref\george{G. Papadopoulos and P.K. Townsend, {\it  Intersecting M-branes},
DAMTP preprint R/96/12, hep-th/9603087.}
\lref\guven{R. G\"uven, Phys. Lett. {\bf 276B} (1992) 49.}
\lref\ght{G. W. Gibbons, G. T. Horowitz and P. K. Townsend, Class. Quantum
Grav. {\bf 12} (1995) 297, hep-th/9410073.}
\lref\jhs{J. H. Schwarz, Phys. Lett. {B360} (1995) 13,
ERRATUM-ibid. {\bf B364} (1995) 252,
 hep-th/9508143.}
\lref\dghw{A. Dabholkar, J. P. Gauntlett, J. A. Harvey and D. Waldram,
{\it Strings as Solitons \& Black Holes as Strings}, hep-th/9511053.}
\lref\callan{C. G. Callan Jr., J. M. Maldacena, A. W. Peet,
{\it Extremal Black Holes as Fundamental Strings}, hep-th/9510134.}
\lref\garf{D. Garfinkle, Phys. Rev. {\bf D46} (1992) 4286.}
\lref\khuri{R. Khuri, Phys. Rev. {\bf D48} (1993) 2947, hep-th/9305143.}
\lref\kounnas{C. Kounnas,  {\it Construction of String Solutions Around
Nontrivial Backgrounds},
in Proceedings of
``From Superstrings to Supergravity", Erice, Italy, Dec 5-12, 1992,
World Scientific, hep-th/9302058.}
\lref\dbranenotes{J. Polchinski, S. Chaudhuri and C.V. Johnson,
{\it Notes on D-Branes}, preprint NSF-ITP-003, hep-th/9602052.}
\lref\green{M.B. Green and M. Gutperle, {\it Light-Cone Supersymmetry and
D-Branes}, preprint DAMTP/96-34, hep-th/9604091.}
\lref\douglas{M . R. Douglas, {\it Branes Within Branes}, Rutgers University
preprint RU-95-92, hep-th/9512077.}
\lref\tseytlin{A.A. Tseytlin, {\it Harmonic Superpositions of M-branes},
Imperial College preprint Imperial/TP/95-95/38, hep-th/9604035.}
\lref\cvetic{M. Cvetic and D. Youm, Phys. Rev. {\bf D53} (1996) 584, 
hep-th/9507090.}
\lref\lifschytz{G. Lifschytz,
{\it Comparing $D$-Branes to Black-Branes}, preprint BRX-TH-394,
hep-th/9604156.}
\lref\rahmfeld{J. Rahmfeld,
Phys. Lett. {\bf B372} (1996) 198,
hep-th/9512089}
\lref\bdgpt{E. Bergshoeff, M de Roo, M.B. Green, G. Papadopoulos and P.K.
Townsend,
{\it Duality of Type II 7-Branes and 8-Branes}, hep-th/9601150 .}
\lref\bho{E. Bergshoeff, C. Hull and T. Ortin,
Nucl. Phys. {\bf B451} (1995) 547,
hep-th/9504081 .}
\lref\bderoo{E. Bergshoeff and M. de Roo, {\it D-Branes and T-Duality},
preprint UG-1/96,
hep-th/9603123}
\lref\sen{A. Sen,
Nucl. Phys. {\bf B434} (1995) 179, hep-th/9408083.}
\lref\dilatonbh{G.W. Gibbons, Nucl. Phys. {\bf B207} (1982) 337;
G.W. Gibbons and K. Maeda, Nucl. Phys. {\bf B298} (1988) 741;
D. Garfinkle, G.T. Horowitz and A. Strominger, Phys. Rev. {\bf D43} (1991)
3140.}
\lref\win{K.Z. Win, {\it Ricci Tensor of Diagonal Metric}, UMASS preprint
gr-qc/9602015.}
\lref\klebanov{I.R. Klebanov and A.A. Tseytlin, {\it Intersecting $M$-branes as
Four-Dimensional Black Holes}, preprint PUPT-1616, Imperial/TP/95-96/41, 
hep-th/9604166 .}
\lref\bala{V. Balasubramanian and F. Larsen, {\it On $D$-Branes and Black Holes in Four Dimensions},
preprint PUPT-1617, hep-th/9604189.}
\lref\calmal{C.G. Callan, Jr. and J. M. Maldacena, {\it $D$-Brane Approach to
Black Hole Quantum Mechanics}, preprint PUPT-1591, hep-th/9602043.}
\lref\ct{M. Cvetic and A.A. Tseytlin, 
Phys. Rev. {\bf D53} (1996) 5619, hep-th/9512031;
A.A. Tseytlin, {\it Dyonic Black Holes in String
Theory}, preprint Imperial-TP-95-96-22, hep-th/9601177.}


\Title{\vbox{\baselineskip12pt
\hbox{CALT-68-2055,
QMW-PH-96-8}
\hbox{UMHEP-429, hep-th/9604179}}}
{\vbox{\centerline{\titlerm  Overlapping Branes in $M$-Theory}
 }}
{\baselineskip=12pt
\centerline{Jerome P. Gauntlett$^{a}$, David A. Kastor$^{b,1}$
and Jennie Traschen$^{b,2}$  }
\medskip
\centerline{\sl ${}^a$ California Institute of Technology}
\centerline{\sl Lauritsen Laboratory, Pasadena, CA 91125}
\centerline{\it and}
\centerline{\sl Department of Physics}
\centerline{\sl Queen Mary and Westfield College}
\centerline{\sl Mile End Road, London E1 4NS, U.K.}
\centerline{\it jerome@theory.caltech.edu}
\medskip
\centerline{\sl ${}^{b}$ Department of Physics and Astronomy}
\centerline{\sl University of Massachusetts}
\centerline{\sl Amherst, MA 01003-4525}
\centerline{\it ${}^{1}$kastor@phast.umass.edu,
${}^{2}$lboo@phast.umass.edu}}
\medskip
\centerline{\bf Abstract}
We construct new supersymmetric solutions of $D$=11 supergravity describing
$n$ orthogonally ``overlapping" membranes and fivebranes for $n$=2,\dots,8.
Overlapping branes arise after separating intersecting branes in a direction
transverse to all of the branes. The solutions, which generalize known
intersecting brane solutions, preserve at least $2^{-n}$ of the supersymmetry. 
Each pairwise overlap involves a membrane overlapping a membrane in a 0-brane,
a fivebrane overlapping a fivebrane in a 3-brane or a membrane overlapping a
fivebrane in a string. After reducing $n$ overlapping membranes to obtain
$n$ overlapping $D$-2-branes in $D$=10, $T$-duality generates new overlapping
$D$-brane solutions in type IIA and type IIB string theory. Uplifting certain
type IIA solutions leads to the $D$=11 solutions. Some of the new 
solutions reduce to dilaton black holes in $D$=4. Additionally, we present
a $D$=10 solution that describes two $D$-5-branes overlapping in a string.
$T$-duality then generates further $D$=10 solutions and uplifting one of the
type IIA  solutions gives a new $D$=11 solution describing two fivebranes
overlapping in a string.

\Date{April 1996}
\vfill
\eject

\newsec{Introduction}

It has been conjectured that there exists a consistent quantum
theory in $D$=11 whose low-energy limit is
$D$=11 supergravity \mtheory,\mtheoryt.
The so called ``$M$-theory" enables one to understand
various duality properties of string theory in a unified way.
The membrane and fivebrane solutions of $D$=11 supergravity preserve
half of the supersymmetry and have played an important role in
elucidating the properties of $M$-theory.

It is natural to enquire whether
there are more general supersymmetric configurations involving
membranes and fivebranes.
Papadopoulos and Townsend \george\ have
recently
re-interpreted certain solutions of $D$=11 supergravity \guven , which preserve
only
$2^{-n}$ of the supersymmetry with $n$=2,3, as describing
$n$ orthogonal membranes intersecting at a point.
Dual solutions were also given in \george\ describing, for $n$=2,
a pair of fivebranes intersecting on a $3$-brane and, for $n$=3,
three fivebranes
intersecting pairwise on $3$-branes, which in turn intersect on a string.

In this paper, we present generalizations of the solutions of \george,\guven\
which preserve at least $2^{-n}$ of the supersymmetry with $n=2,\dots,8$ and
include $n$ arbitrary harmonic functions. For $n=2,3$ the solutions
preserve exactly $2^{-n}$ of the supersymmetry. These new solutions
describe branes which 
``overlap" orthogonally in a certain number of tangent
directions: overlapping branes
are obtained from intersecting branes by separating one
of the branes in a direction transverse to all of the branes.
Branes intersect only if two or more of the $n$
functions are singular at the same point.
For $n=2,3$ the solutions of \george,\guven\ 
are recovered when the $n$ harmonic functions
are
taken to be equal. Included in our new solutions are 
membranes and fivebranes overlapping in one tangent direction;
in the intersecting case, $D$=11 membranes and fivebranes are found which
intersect on a
string\foot{While this work was
being completed, a
paper by Tseytlin \tseytlin\ appeared which contains many of the results
presented in sections (2) - (4) below.}.

We begin in section (2) by giving overlapping
membrane solutions with $n$=2,3,4.  We then show in section (3) that other
$D$=11 solutions, involving fivebranes and collections of membranes and
fivebranes all preserving $2^{-n}$ of the supersymmetry, 
can be obtained from the membrane solutions of section (2)
via dimensional reduction, performing $T$-duality transformations and then
uplifting back to $D$=11.
While carrying out these transformations, we will also find solutions in lower
dimensions which describe $n$ overlapping $D$-branes.
Recall that all single, {\it i.e.}, $n$=1,
$D$-brane solutions may be obtained starting from any given one, {\it e.g.},
the $D$-$0$-brane,
by acting with $T$-duality \bdgpt,\bderoo . It seems natural to
conjecture that a similar relation holds between all solutions involving $n$
overlapping $D$-branes, which preserve the same amount of the supersymmetry.
We also discuss solutions with $n\ge 4$, which preserve more than $2^{-n}$ of the 
supersymmetry due to the presence of certain special triple overlaps.
In section (4), we will discuss the reduction
of the $n$=1,2,3 overlapping brane solutions to dilaton black holes solutions
in
$D$=4.  We find, for example, that the overlapping membrane (fivebrane)
solutions reduce to 
black  holes carrying $n$ types of electric (magnetic) charge.
The $n$=2 solutions belong to a class recently studied in \cvetic,\rahmfeld .
Further,  we
find that all the intersecting cases reduce to extreme
dilaton black holes with $a=\sqrt{(4/n)-1}$, as for
the cases considered in \george.  We discuss how this comes about at the level
of the
dimensionally reduced action.

It was pointed out in \george\ that the solutions of \george,\guven\
are not ``true" $D$=11 intersections in the sense that the harmonic function
corresponding to a given brane is translationally invariant
in the directions tangent to all of the branes. This property is
shared by the generalized overlapping solutions of sections (2)-(4).
In section (5) we will
interpret a solution first constructed by Khuri that preserves $1/4$ of
the supersymmetry \khuri\ as describing a true overlap of two
5-branes in a string in $D$=10 type II supergravity.
New solutions can be generated from this using $SL(2,Z)$ and $T$-duality.
By uplifting to $D$=11 we obtain a solution describing two fivebranes
overlapping in a string, although it is not a true overlap.
Section (6) concludes with some comments.

\newsec{Overlapping Membranes}
The bosonic field content of $D$=11 supergravity consists of a metric
and a three-form potential with four-form field 
strength\foot{All $D$=11 fields have tilde's.
All $D$=10, IIA supergravity fields will
have hats and $D$=10, IIB fields will have bars.}
$\tilde F_{MNPQ}=24\nabla_{[M}\tilde{A}_{NPQ]}$.
The action for the bosonic fields is given by
\eqn\action{S=\int\sqrt{-\tilde{g}}\left\{
R-{1\over 12}\tilde{F}^2-{1\over 10,368}
\epsilon^{\mu_1\dots\mu_{11}}\tilde{F}_{\mu_1\dots\mu_4}
\tilde{F}_{\mu_5\dots\mu_8}
\tilde{A}_{\mu_9\dots\mu_{11}}\right\}.
}
Supersymmetric solutions to the corresponding equations of motion
can be constructed
by looking for bosonic backgrounds that admit Killing spinors {\it i.e.},
backgrounds which admit spinors such that the supersymmetry
variation of the gravitino field $\tilde\psi_M$ vanishes:
\eqn\susy{
\left[D_M+
{1\over 144}({\Gamma_M}^{NPQR}-8\delta^N_M\Gamma^{PQR})
\tilde F_{NPQR}\right]\epsilon
=0,
}
where $\epsilon$ is a 32-component Majorana spinor.

We begin by presenting the generalized supersymmetric solution
for $n$=2 orthogonally overlapping
membranes\foot{We will often refer to both membranes and anti-membranes as
membranes.},
\eqn\twomembranes{\eqalign{
d\tilde{s}^2 = &-\left(f_1f_2\right)^{-2/3}dt^2 +
f_1^{-2/3}f_2^{1/3}\left(dx_1^2+dx_2^2\right) +
f_1^{1/3}f_2^{-2/3}\left(dx_3^2+dx_4^2\right)\cr &+
\left(f_1f_2\right)^{1/3}\left(dx_5^2+\dots +dx_{10}^2\right),\cr
&\tilde{F}_{t12\alpha} = {c_1\over 2}{\partial_\alpha f_1\over f_1^2}, \qquad
\tilde{F}_{t34\alpha} = {c_2\over 2}
{\partial_\alpha f_2\over f_2^2}, \qquad \a=5,\dots, 10.\cr
&f_i= f_i(x_5,\dots ,x_{10}), \qquad \nabla^2f_i=0,\qquad c_i=\pm1,\qquad
i=1,2.\cr
}}
It is straightforward to show that this background preserves $1/4$
of the supersymmetry: there are eight Killing spinors of the form
$\epsilon=(f_1 f_2)^{-1/6}\eta$, where $\eta$ is constant and satisfies
the algebraic constraints
\eqn\con{
\eqalign{\hat\Gamma^0\hat\Gamma^1\hat\Gamma^2\eta&=c_1\eta\cr
\hat\Gamma^0\hat\Gamma^3\hat\Gamma^4\eta&=c_2\eta,\cr}
}
where $\hat\Gamma^M$ are Gamma matrices in an orthonormal frame.
These conditions can be recast, as in \guven , in the compact form
\eqn\alg{
\hat \Gamma_a\eta=\hat\Gamma_0{J_a}^b\hat\Gamma_b\eta,
}
where the indices $a,b=1,\dots4$ correspond to the
membrane coordinates $x^1,\dots x^4$
and ${J_a}^b$ is a complex structure on this space (depending
on the $c_i$).

The functions $f_i$ are harmonic in the coordinates
$\vec x=\{x_5,\dots,x_{10}\}$ and we first take them to be of the form
\eqn\harmonica{
f_i = 1 + {M_{i}\over r_{i}^4},\qquad
r_{i}= \left |\vec{x}-\vec{x}_{i}\right |. }
The solution then describes a membrane oriented in the $(1,2)$ plane with
position $\vec{x}_{1}$ and another oriented in the $(3,4)$ plane with
position $\vec{x}_2$ orthogonally overlapping in a point. 
In particular note that the directions tangent
to the $i$th set of membranes appear with the
power
$f_i^{-2/3}$  and those transverse with the power $f_i^{1/3}$, which are the
appropriate
powers for the $n$=1 membrane  solutions.
A
membrane with $(1,2)$ orientation intersects one with $(3,4)$ orientation in
the degenerate case that
$\vec{x}_{1}=\vec{x}_{2}$. 
Note that the $n$=2 solutions of G\"uven \guven\ are 
recovered by taking $f_1=f_2$.
These observations clarify the
interpretation \george\ of the G\"uven solutions as an intersection of two
membranes in the
degenerate limit.

The most general solution has harmonic functions of the form
\eqn\harmonicb{
f_i = 1 + \sum_{I=1}^{k_i}{M_{i,I}\over r_{i,I}^4},\qquad
r_{i,I}= \left |\vec{x}-\vec{x}_{i,I}\right |. }
The solution then 
describes $k_1$ parallel membranes with $(1,2)$ orientation and
positions $\vec{x}_{1,I}$, and $k_2$ parallel membranes with $(3,4)$ 
orientation and
positions $\vec{x}_{2,I}$. Each membrane of one set orthogonally overlaps
all of the membranes in the other set in a point.
A membrane with $(1,2)$ orientation intersects one with $(3,4)$ orientation in
the degenerate case that
$\vec{x}_{1,I}=\vec{x}_{2,J}$, for some combination $I,J$.
Note that in describing the solutions in the rest 
of the paper we will implicitly
take the harmonic functions to be that of a single brane as in
\harmonica\ for ease of exposition.

Like the intersecting membrane solutions of \george,\guven , these new
solutions have the
property that the metric is invariant not only under translations in
common tangent directions\foot{We will generally follow the terminology of
\george\ in
referring to
{\it common tangent} directions as being tangent directions common
to all
branes. {\it Relative transverse} directions are those tangent to 
at least one but not
all 
branes and {\it overall transverse} directions are those orthogonal to all
branes.},
{\it i.e.}, the time direction in this case, but also
under translations in all the relative transverse directions.
For example, $f_1$ does not fall off in the $x_3,x_4$ directions, as one would
expect for
a $D$=11 membrane spatially oriented in the $(1,2)$ plane.
This suggests that, as in \george , the solutions \twomembranes\ be considered
in a
dimensionally reduced context, with all relative transverse directions
periodically identified. This implies, {\it e.g.}, that
the membrane with spatial orientation in the $(1,2)$ plane has been wrapped in
these
directions and directly
reduced in the $(3,4)$ directions.
The solutions \twomembranes\ may then be regarded as $D$=7 black
holes.
The black holes would each, generically, carry one of two types of electric
charge,
depending on the $10$-dimensional orientation of the corresponding membrane.
In the
degenerate intersecting case \george , black holes would carry equal amounts
of both types of electric charge.

The solutions \twomembranes\ are generically singular on the surfaces
$\vec{x}-\vec{x}_{i}=0$, with the scalar curvature diverging.
This behavior is different from the $n$=1 membrane case, where these
surfaces are
regular event horizons \ght.
The singularity in the present case arises because the
functions
$f_i$ are constant in the relative transverse dimensions, rather than having
the fall-off characteristic of a single membrane.  Interestingly, in the
intersecting case, $f_1=f_2$, the scalar curvature is finite and this 
singularity is removed. The singularity reappears if the intersecting solutions
are further dimensionally reduced in overall transverse directions.

The generalization to $n$=1,2,3,4 membranes may be written in a unified fashion
as
\eqn\alln{\eqalign{
d\tilde{s}^2 = &-\prod_{i=1}^n f_i^{-2/3}dt^2
+ \sum_{i=1}^n \left[f_i^{-2/3}\prod_{j\neq i}f_j^{1/3}\right]
\left(dx_{2i-1}^2+dx_{2i}^2\right) +
\cr &
+ \prod_{i=1}^nf_i^{1/3}
\left(dx_{2n+1}^2+\dots +dx_{10}^2\right)\cr
\tilde{F}_{t12\a} = & {c_1\over 2}{\partial_\a f_1\over f_1^2}, \dots,
\tilde{F}_{t,2n-1,2n,\a} = {c_n\over 2}
{\partial_\a f_n\over f_n^2}, \qquad \a=2n+1,
\dots, 10.\cr
f_i=& f_i(x_{2n+1},\dots ,x_{10}), \qquad \nabla^2f_i=0,\qquad c_i=\pm 1,
\qquad i=1,\dots,n.\cr
}}
These solutions preserve $2^{-n}$ of the supersymmetries. The Killing
spinors are given by $\epsilon=\prod_{i=1}^nf_i^{-1/6}\eta$ where the
constant spinor $\eta$ satisfies the obvious algebraic constraints generalizing
\con, which can again be recast in the form \alg\ with $J$ a complex
structure on the relative transverse space of the membranes.

The $n$=3 solutions describe three membranes with orientations in the
$(1,2)$, $(3,4)$ and $(5,6)$ directions respectively.  If we consider these
relative
transverse dimensions to be compact, we get \george\ $n$=3
$D$=5 black holes, each
carrying,
generically, one of three different types of electric charge.  The $n$=4
solutions give
$D$=3 objects, each carrying one of four types of electric charge.
In this case the solutions have bad asymptotics, the functions $f_i$ diverging
logarithmically.  However, they share many formal properties with the $n<4$
solutions, including (a lesser amount of) unbroken supersymmetry, and Sen \sen\
has made use of similar objects in his study of strong-weak coupling
duality in three-dimensional string theory.
We note that the components of the stress tensor in an orthonormal frame
fall off faster than $1/r$ for these objects.  One may
also consider $n$=5, however, in this case, the total transverse space has
shrunk
to zero dimensions and
the functions $f_i$ are constant, giving $D$=11 Minkowski space.

\newsec{New Overlapping Branes From Old Via $T$-Duality}
\subsec{Dimensional Reduction and Type II $T$-Duality Rules}
Rather than present the new $D$=11 solutions
involving fivebranes in a similar direct
fashion,
in this section we will show how they may be obtained from the membrane
solutions \alln,
making use of dimensional reduction and $T$-duality. In particular, starting
from the $n$=2
membrane  solution, we will obtain a fivebrane overlapping a fivebrane on a
$3$-brane and also a fivebrane and a membrane 
overlapping on a string.  Starting from
the $n$=3 membrane solution, we will find solutions with three 
fivebranes, solutions
with a membrane and two fivebranes and solutions with a fivebrane and
two membranes. Similar overlapping solutions may be constructed from the
$n$=4 membrane
solution, but we will not write them down explicitly. However, we will explain
why for $n\ge 4$ there
are also overlapping solutions that preserve more than $2^{-n}$ of the
supersymmetry.
In addition to these $D$=11 solutions we
will also see how one can construct various overlapping brane solutions
of type IIA and type IIB string theory in $D$=10.

There are two distinct ways to
dimensionally reduce the membrane solutions \alln\ to get solutions of $D$=10,
IIA
supergravity.  We will choose to reduce along one of the overall transverse
directions, taking the functions $f_i$ to be independent of, {\it e.g.},
$x_{10}$.  This gives 
$n$ membranes overlapping in a point
in $D$=10, which are $D$-2-branes of the IIA theory.
If instead one reduces
along one of the relative transverse directions one obtains a 
fundamental string and $n$-1 $D$-2-branes overlapping in a point.
For present purposes, we need only a limited form
of the
relations between the fields  of $D$=11 supergravity and those of
$D$=10, IIA supergravity,
given by\foot{A complete derivation of the reduction from $D$=11 to
$D$=10, IIA and the Type II $T$-duality rules is given in \bho .}
\eqn\eleventoten{
\tilde{g}_{10,10}=e^{4\hphi /3},\qquad
\tilde{g}_{\mu\nu}=e^{-2\hphi /3}\hat{g}_{\mu\nu},\qquad
\tilde{F}_{\mu\nu\rho\sigma}=\hat{F}_{\mu\nu\rho\sigma},}
where $\mu,\nu =t,1,\dots, 9$ and
we have assumed that $\tilde{g}_{\mu,10}=F_{10,\mu,\nu,\rho}=0$.
The $n$=2 membrane solution \twomembranes\ then reduces along $x_{10}$ to give
\eqn\tendmembranes{\eqalign{
d\hat{s}^2= &-\left(f_1f_2\right)^{-1/2}dt^2 +
f_1^{-1/2}f_2^{1/2}\left(dx_1^2+dx_2^2\right) +
f_1^{1/2}f_2^{-1/2}\left(dx_3^2+dx_4^2\right)\cr &+
\left(f_1f_2\right)^{1/2}\left(dx_5^2+\dots +dx_{9}^2\right),\qquad
f_i=f_i(x_5,\dots,x_9),
\cr
e^{-2\hphi}= & \left(f_1f_2\right)^{-1/2},\qquad
\hat{F}_{t12\a} = {c_1\over 2}{\partial_\a f_1\over f_1^2}, \qquad
\hat{F}_{t34\a} = {c_2\over 2}{\partial_\a f_2\over f_2^2}.\cr }}

We can now use type II $T$-duality to find new overlapping brane solutions.
Recall that type II $T$-duality is a map from solutions of IIA supergravity
into solutions of
IIB supergravity, and vice-versa.
For diagonal metrics, the action of type II $T$-duality, with respect to a
symmetry direction $\alpha$, on
the dilaton and metric is simply given by \bho,
\eqn\tdualrules{\bar{g}_{\alpha\alpha}=1/\hat{g}_{\alpha\alpha},\qquad
e^{-2\bar{\varphi}}=
\hat{g}_{\alpha\alpha}e^{-2\hat{\phi}}, }
where $\hphi$ is the IIA dilaton and $\bar{\varphi}$ is
the IIB dilaton.  The action of $T$-duality on the gauge fields is more
involved \bho\
and we display only the final results below.

There are two different ways to act with
$T$-duality on \tendmembranes .
Acting on one of the overall transverse directions maps it
into a common tangent direction, increasing the dimension of both 
branes by one.
In this way, we can get two $D$-3-branes of type IIB
overlapping on a string, two 
$D$-4-branes of type IIA
overlapping on
a membrane and so on.   
Alternatively, we can act with $T$-duality on one of the
relative  transverse dimensions, increasing the dimension of one brane
and decreasing the
dimension of the other.  In this way, we can construct a $D$-string 
overlapping a 
$D$-3-brane of type IIB 
in a point and a $D$-0-brane overlapping a $D$-4-brane of type IIA in a point.  
By
combining both
sorts of
transformations we can construct a variety of overlapping brane
solutions which we record here:
\eqn\list{
\eqalign{IIA:\qquad &0\cap 4 (0);\quad 2\cap 2 (0);\quad 2\cap 4 (1);
\quad 2\cap 6 (2);
\quad 4\cap 4 (2);\cr
&4\cap6 (3);\quad 4\cap 8 (4); \quad 6\cap6 (4);
\quad 6\cap 8 (5)\cr
IIB:\qquad &1\cap 3 (0);\quad 1\cap 5 (1);
\quad  3\cap 3 (1);\quad  3\cap 5 (2);\quad  3\cap 7 (3);\cr
& 5\cap 5 (3);\quad  5\cap 7 (4);
\quad  5\cap 9 (5);\quad 7\cap 7 (5);
\cr}
}
where $3\cap 5 (2)$ denotes a $D$-3-brane overlapping
a $D$-5-brane in a membrane and so on.

Let us pause to make a few comments on this
table. Firstly, to implement the $T$-duality
transformations properly 
for $p\ge7$ one must use the recently formulated massive type
II
supergravity \bdgpt. Secondly, we note that
acting with $SL(2,Z)$ duality will give new solutions describing
overlapping branes carrying R-R and NS-NS charges.
Thirdly, the solution $1\cap5(1)$ was constructed in \calmal\ (see also
\ct) in the
context of determining black hole entropy using $D$-brane techniques.
Finally, note that for the cases where the branes overlap in a 5-brane
the overall transverse directions have shrunk to a point and the solution is
just
Minkowski space. However, in the case $5\cap9(5)$, since at the
level of classical solutions a $D$-9-brane is
just Minkowski space, this overlap is just a $D$-5-brane which obviously
does correspond to a non-trivial
solution. This suggests that
non-trivial overlapping solutions may also exist
in the other cases in this class, possibly of the type discussed in section 5.

We can compare the list \list\ of overlapping $D$-brane field theory
solutions with supersymmetric
superpositions of $D$-branes allowed at the string theory level.
Recall that open strings have Neumann (N) boundary conditions in directions
tangent to a $D$-brane and Dirichlet (D) boundary conditions in the transverse
directions.  In a configuration of overlapping $D$-branes, certain open string
coordinates may have Dirichlet boundary conditions at one end of the string and
Neumann boundary conditions at the other end.  It is argued in \dbranenotes ,
that in order for a superposition of $D$-branes to have some amount of unbroken
supersymmetry, the number of these ND directions must be a multiple of four 
\foot{This condition can also be arrived at by determining which
$D$-brane configurations satisfy a zero force condition
\green,\lifschytz. The special case of ``embedded" branes, where the
dimension of the intersection is the same as one of the branes, has
been studied by Douglas \douglas.}.
In our solutions, the number of ND directions is equal to the size of the
relative transverse space, which has dimension four for all the solutions in
\list\ and one can check that this is a complete list involving
two branes.
Solutions with eight ND directions will be presented in section (5).

The main purpose of this paper is to construct supersymmetric
solutions of $M$-theory describing overlapping
fivebranes and membranes. This can be achieved by uplifting the type IIA
overlapping solutions involving $D$-4-branes and $D$-2-branes and the
results are presented in the next subsections. Uplifting
the other type IIA branes to $D$=11 is left for future work.

\subsec{Fivebrane and Fivebrane Overlap}
The following series of steps yields overlapping fivebranes in $D$=11.
Taking both functions $f_i$ in \tendmembranes\ to be independent of $x_5$ and
$x_6$,  act
successively  with T-duality in these two directions. This gives a IIA solution
with two 
$D$-4-branes, oriented in the $(1256)$ and $(3456)$
hyperplanes, overlapping on
a membrane in
the $(56)$ plane, with metric and dilaton given by
\eqn\fourbranes{\eqalign{d\hat{s}^2=
&\left(f_1f_2\right)^{-1/2}(-dt^2+dx_5^2+dx_6^2)
+  f_1^{-1/2}f_2^{1/2}\left(dx_1^2+dx_2^2\right)+
f_1^{1/2}f_2^{-1/2}\left(dx_3^2+dx_4^2\right)\cr &+
\left(f_1f_2\right)^{1/2}\left(dx_7^2+dx_8^2+dx_{9}^2\right),\qquad
e^{-2\hphi}= \left(f_1f_2\right)^{1/2}.\cr }}
Lifting back to $D$=11, the dilaton turns into a common tangent direction and
we get
\eqn\fivebranes{\eqalign{
d\tilde{s}^2 = &\left(f_1f_2\right)^{-1/3}(-dt^2 +dx_5^2 +dx_6^2
+dx_{10}^2)+
f_1^{-1/3}f_2^{2/3}\left(dx_1^2+dx_2^2\right) \cr & +
f_1^{2/3}f_2^{-1/3}\left(dx_3^2+dx_4^2\right)+
\left(f_1f_2\right)^{2/3}\left(dx_7^2+dx_8^2+ dx_9^2\right). \cr
\tilde{F}_{34\a\b}= &{c_1\over 2}\epsilon_{\a\b\g}\partial_\g f_1,\qquad
\tilde{F}_{12\a\b}= {c_2\over 2}\epsilon_{\a\b\g}\partial_\g f_2,
\qquad f_i=f_i(x_7,x_8,x_9),\cr}}
where $\epsilon_{\a\b\g}$ is the $D$=3 flat space alternating symbol.
The solution preserves $1/4$ of the supersymmetry and the Killing
spinors are given by $\epsilon=(f_1 f_2)^{-1/12}\eta$ with the constant
spinor $\eta$ satisfying the constraints
\eqn\ffcon{
\eqalign{
&\hat\Gamma^3\hat\Gamma^4\gamma^*\eta=c_1\eta\cr
&\hat\Gamma^1\hat\Gamma^2\gamma^*\eta=c_2\eta,\cr}
}
with $\gamma^*=\hat\Gamma^7\hat\Gamma^8\hat\Gamma^9$. This
can also be recast, as in \george\ in the form
\eqn\ffcs{
\hat\Gamma_a\eta={J_a}^b\hat\Gamma_b\gamma^*\eta,}
where $J$ is again the appropriate complex structure on the relative
transverse space of the fivebranes.

This solution describes two fivebranes, oriented in the $(1,2,5,6,10)$
and
$(3,4,5,6,10)$ hyperplanes, overlapping on a 3-brane in the $(5,6,10)$
hyperplane.
Setting $f_1=f_2$, it
reduces to the intersecting fivebrane solution of \george .
Reducing \fivebranes\ along all four relative transverse dimensions gives, as in
\george , 
$D$=7 3-branes, which are magnetic duals to the $D$=7 black
holes,
which came from
reducing the $n$=2 membrane solution \twomembranes .  As with \twomembranes ,
the solution
\fivebranes\ is singular at the positions of the branes due to the lack of
fall-off
of the functions $f_i$ in the relative transverse directions.  Unlike in
the case
of $n$=2 membranes, this singularity is not removed in the intersecting case.

\subsec{Membrane and Fivebrane Overlap}
We get $n$=2 solutions with a membrane  and a fivebrane
in $D$=11 by a slightly
different
path.  Starting again with \tendmembranes ,  T-dualize on one dimension
parallel to
a membrane, {\it e.g.} $x_3$, to get a IIB solution describing a
string overlapping a 3-brane in a point,
\eqn\onethree{\eqalign{
d\bar{s}^2= &\left(f_1f_2\right)^{-1/2}(-dt^2) +
f_1^{-1/2}f_2^{1/2}\left(dx_1^2+dx_2^2+dx_3^2\right) +
f_1^{1/2}f_2^{-1/2}\left(dx_4^2\right)\cr &+
\left(f_1f_2\right)^{1/2}
\left(dx_5^2+dx_6^2+dx_7^2+dx_8^2+dx_{9}^2\right),\qquad
e^{-2\bar{\varphi}}= f_2^{-1}.\cr }}
Next, take both functions $f_i$ to be independent of the coordinate $x_5$ and
$T$-dualize in
that direction to give a IIA solution describing a $D$-4-brane overlapping a
membrane in a
string,
\eqn\membranefourbrane{\eqalign{
d\hat{s}^2= &\left(f_1f_2\right)^{-1/2}(-dt^2+dx_5^2) +
f_1^{-1/2}f_2^{1/2}\left(dx_1^2+dx_2^2+dx_3^2\right) +
f_1^{1/2}f_2^{-1/2}\left(dx_4^2\right)\cr &+
\left(f_1f_2\right)^{1/2}\left(dx_6^2+dx_7^2+dx_8^2+dx_{9}^2\right),\qquad
e^{-2\hphi}= f_1^{1/2}f_2^{-1/2}.\cr }}
Finally, lift back to $D$=11 to get a fivebrane overlapping a membrane in a
string
\eqn\membranefivebrane{\eqalign{
d\tilde{s}^2 = &f_1^{-1/3} f_2^{-2/3}(-dt^2 +dx_5^2)+
f_1^{-1/3}f_2^{1/3}\left(dx_1^2+dx_2^2+dx_3^2+dx_{10}^2\right)\cr & +
f_1^{2/3}f_2^{-2/3}\left(dx_4^2\right)+
f_1^{2/3}f_2^{1/3}\left( dx_6^2+dx_7^2+dx_8^2+dx_9^2\right),\cr
\tilde{F}_{4\a\b\g}=&{c_1\over 2}\epsilon_{\a\b\g\d}\partial_\d f_1,\qquad
\tilde{F}_{t45\a}={c_2\over 2}{\partial_\a f_2\over f_2^2},\qquad
f_i=f_i(x_6,\dots,x_9),}}
where $\epsilon_{\a\b\g\d}$ is the $D$=4 flat space alternating symbol.
The eight Killing spinors have the form $\epsilon=f_1^{-1/12}f_2^{-1/6}\eta$
with the constant spinor $\eta$ satisfying
\eqn\tfcon{
\eqalign{
&\hat\Gamma^0\hat\Gamma^4\hat\Gamma^5\eta=c_1\eta\cr
&\hat\Gamma^4\hat\Gamma^6\hat\Gamma^7\hat\Gamma^8\hat\Gamma^9\eta=c_2\eta.\cr}
}
This solution describes a membrane in the $(4,5)$ direction and a fivebrane
in the $(1,2,3,5,10)$ direction, overlapping in a string in the $(5)$
direction.
Reducing \membranefivebrane\ along all relative transverse dimensions gives
electric and magnetic strings in $D$=6.  Electric-magnetic
duality
then maps within this class of solutions, as expected based on the $D$=11
duality between
membranes and fivebranes .

\subsec{Fivebrane, Fivebrane and Fivebrane Overlap}
We now turn to $n$=3 solutions involving fivebranes.
Begin by reducing the
$n$=3 membrane
solution \alln\ to $D$=10, IIA by taking the functions $f_i$ independent of
$x_{10}$,
giving
\eqn\threemembranes{\eqalign{
d\hat{s}^2 = &-\prod_{i=1}^3 f_i^{-1/2}dt^2
+ \sum_{i=1}^3\left[ f_i^{-1/2}\prod_{j\neq i}f_j^{1/2}\right]
\left(dx_{2i-1}^2+dx_{2i}^2\right) 
\cr &+ \prod_{i=1}^3 f_i^{1/2}
\left(dx_{7}^2+dx_8^2+dx_{9}^2\right),\qquad
e^{-2\hphi}= \prod_{i=1}^3 f_i^{-1/2}, \cr
}}
which describes three $D$-2-branes in $D$=10.
Acting with $T$-duality in different ways generates a set of solutions
of type II supergravity, which describe three overlapping
branes, which all preserve $1/8$ of the
supersymmetry.
We will not present this list here but we note that the
intersections of any pair of $D$-branes is necessarily one from \list.
To get intersections of membranes and fivebranes in $M$-theory we again 
want to uplift the solutions of type IIA supergravity involving $D$-2-branes
and $D$-4-branes. The uplifted $D=11$ solutions involving $n=3$ branes
also preserve $1/8$ of the supersymmetry.

If we
act with $T$-duality on \threemembranes\
successively in all
six relative transverse dimensions to get a IIA solution and then lift back to
$D$=11, we get
\eqn\threefivebranes{\eqalign{
d\hat{s}^2 = &\prod_{i=1}^3 f_i^{-1/3}(-dt^2+dx_{10}^2)
+ \sum_{i=1}^3\left[ f_i^{2/3}\prod_{j\neq i}f_j^{-1/3}\right]
\left(dx_{2i-1}^2+dx_{2i}^2\right) 
\cr + &\prod_{i=1}^3 f_i^{2/3}
\left(dx_{7}^2+dx_8^2+dx_{9}^2\right),\cr
\tilde{F}_{12\a\b}=&{c_1\over 2}\epsilon_{\a\b\g}\partial_\g f_1,\qquad
\tilde{F}_{34\a\b}={c_2\over 2}\epsilon_{\a\b\g}\partial_\g f_2,\qquad
\tilde{F}_{56\a\b}={c_3\over 2}\epsilon_{\a\b\g}\partial_\g f_3.\cr
}}
This now describes three fivebranes oriented in the
$(3,4,5,6,10)$,
$(1,2,5,6,10)$ and $(1,2,3,4,10)$ hyperplanes respectively.
Each pair of fivebranes overlaps on a $3$-brane.
The $3$-branes finally all overlap on a string.
Reducing these solutions along all
relative transverse dimensions gives \george\ strings in $D$=5
carrying
three types of magnetic charge, which are dual to the $D$=5 electric black
holes coming
from \alln\ with $n$=3.

There are two additional solutions corresponding to three overlapping
fivebranes.
The first may be obtained from \threemembranes\ by
first performing $T$-duality
in one overall transverse direction and on one direction each from the
membranes,  {\it e.g.}, the $x_7, x_2,x_4$ and $x_6$ directions. Uplifting
this solution to $D$=11 gives fivebranes in the $(1,4,6,7,10)$,
$(2,3,6,7,10)$ and
$(2,4,5,7,10)$ directions. Pairwise the fivebranes overlap in 3-branes 
and the latter overlap in a 2-brane. For this solution there are
two overall transverse directions and the harmonic functions logarithmically
diverge. The second solution can be obtained from \threemembranes\ by
$T$-dualizing on two overall transverse directions, {\it e.g.}, $x_7, x_8$.
Uplifting to $D$=11 gives fivebranes in the $(1,2,7,8,10)$,
$(3,4,7,8,10)$ and
$(5,6,7,8,10)$ directions. In this case the overlap of the pairwise overlaps is
a 3-brane.
The harmonic functions now grow linearly as there is
only one overall transverse direction.

\subsec{Membrane, Fivebrane and Fivebrane Overlap}
Again starting from \threemembranes , act successively with $T$-duality on both
spatial
tangent directions of one $D$-2-brane and one each of the tangent directions
of the two
other branes, 
{\it e.g.}, the $x_1,x_2,x_3$ and $x_5$ directions.  Lifting back to
$D$=11
then gives
\eqn\twofivefive{\eqalign{
d\tilde{s}^2 = &  -f_1^{-2/3}(f_2f_3)^{-1/3}dt^2 +
f_1^{1/3}(f_2f_3)^{-1/3}\left(
dx_1^2 + dx_2^2 +dx_{10}^2\right) \cr &
+f_1^{-2/3}f_2^{2/3}f_3^{-1/3}dx_3^2 +f_1^{1/3}f_2^{-1/3}f_3^{2/3}dx_4^2 \cr &
+f_1^{-2/3}f_2^{-1/3}f_3^{+2/3}dx_5^2 +f_1^{1/3}f_2^{2/3}f_3^{-1/3}dx_6^2\cr &
f_1^{1/3}f_2^{2/3}f_3^{2/3}\left(dx_7^2 + dx_8^2 +dx_9^2\right), \cr
\tilde{F}_{t35\a}=&{c_1\over 2}{\partial_\a f_1\over f_1^2},\qquad
\tilde{F}_{36\a\b}={c_2\over 2}\epsilon_{\a\b\g}\partial_\g f_2,\qquad
\tilde{F}_{45\a\b}={c_3\over 2}\epsilon_{\a\b\g}\partial_\g f_3,\cr
}}
which describes a membrane in the $(3,5)$ plane and two fivebranes in the
$(1,2,4,5,10)$ and
$(1,2,3,6,10)$ hyperplanes.  The fivebranes overlap on a $3$-brane. The
membrane overlaps
each of the fivebranes on a string and the strings overlap each  other and the
3-brane at a
point.  Reducing all the relative transverse directions gives $D$=4 black holes
carrying one
type of electric and two types of magnetic charge.

There exists another solution 
where the pairwise overlaps overlap
in a string. It has only two overall transverse directions so the
harmonic functions diverge logarithmically. To obtain the solution start
with \threemembranes\ and perform $T$-duality on one overall transverse
direction and on one of the directions along one of the membranes, {\it e.g.},
$x_2$ and $x_7$. After uplifting to $D$=11 one finds a membrane in
the $(1,7)$ direction and fivebranes in the $(2,3,4,7,10)$ and
$(2,5,6,7,10)$ directions.

\subsec{Membrane, Membrane and Fivebrane Overlap}
Again, start with \threemembranes .  Act successively with $T$-duality on one
direction each from two of the membranes, {\it e.g.}, $x_1$ and $x_3$, and lift
back to $D$=11 to get
\eqn\twotwofive{\eqalign{
d\tilde{s}^2 = &  -(f_1f_2)^{-2/3}f_3^{-1/3}dt^2 +
f_1^{1/3}f_2^{-2/3}f_3^{-1/3}dx_1^2 +
f_1^{-2/3}f_2^{1/3}f_3^{-1/3}dx_3^2\cr &
+f_1^{-2/3}f_2^{1/3}f_3^{2/3}dx_2^2 +f_1^{1/3}f_2^{-2/3}f_3^{2/3}dx_4^2 \cr &
+f_1^{1/3}f_2^{1/3}f_3^{-1/3}(dx_5^2 +dx_6^2+dx_{10}^2)\cr &
f_1^{1/3}f_2^{1/3}f_3^{2/3}\left(dx_7^2 + dx_8^2 +dx_9^2\right), \cr
\tilde{F}_{t23\a}=&{c_1\over 2}{\partial_\a f_1\over f_1^2},\qquad
\tilde{F}_{t14\a}={c_2\over 2}{\partial_\a f_2\over f_2^2},\qquad
\tilde{F}_{24\a\b}={c_3\over 2}\epsilon_{\a\b\g}\partial_\g f_3. \cr
}}
This describes two membranes in the $(2,3)$ and $(1,4)$ planes and a
fivebrane in
the
$(1,3,5,6,10)$ hyperplane.

\subsec{Four Overlapping Branes or More}
We have not carried out a systematic search for all 
overlapping brane solutions with four or more sets of branes.  
We expect 
that any configuration of branes which satisfies the 
conditions for pairwise overlaps should yield a solution of the sort described 
above.  We would like to call attention, however, to an interesting phenomenon
associated with triple overlaps.  For certain special triple overlaps,
the three gamma matrix projections multiply together to 
give a projection appropriate for another set of branes.
This raises the possibility of adding an additional set of 
branes with this new orientation, without further breaking 
supersymmetry\foot{We assume, for the purposes of this discussion, that such a configuration will be a solution, if it satisfies the pairwise overlap conditions.
We have checked this explicitly in
an example of $n$=5 overlapping fivebranes preserving $1/16$ of the 
supersymmetry, though not in general.}. 

The first special triple intersection is that of two membranes and a 
fivebrane.  Taking the branes to be oriented in the 
$(12)$, $(3,4)$ and $(1,3,5,6,7)$ directions,
the product of the three projections gives the projection for a fivebrane in
the $(2,4,5,6,7)$ hyperplane.  The resulting configuration gives a solution
with $2\cap 2\cap 5\cap 5$ preserving $1/8$ of the 
supersymmetry, which has been 
noted by Klebanov and Tseytlin \klebanov
\foot{Upon dimensional reduction, this solution can be 
related via $T$-duality to {\it e.g.}, a symmetric configuration 
of four overlapping $3$-branes 
\klebanov,\bala. }.  
This solution has three 
overall transverse directions and can be reduced to give a black hole solution
in four dimensions with finite horizon area, 
making it interesting for studies of black hole entropy 
\klebanov,\bala.  Note that the polarization of the fourth set of branes 
({\it i.e.}, whether they are branes or anti-branes) 
is determined by the polarizations of the first
three sets.  

The second special triple overlap can be found from the above case by deleting
one of the membranes.  This gives two fivebranes each overlapping a membrane in
distinct strings.  The deleted membrane can clearly be added back in 
without further breaking supersymmetry.  The third special case involves 
three fivebranes.
As noted above, three fivebranes can overlap either in a common
$3$-brane, a $2$-brane or a string.  In the $2$-brane case, the product of
the three projections gives another fivebrane projection, 
allowing for a solution with four fivebranes and two overall 
transverse directions, which preserves $1/8$ of the supersymmetry. The four
fivebranes overlap in a common 2-brane.
All three special triple overlap
solutions are related by $T$-duality after reduction to IIA $D$-branes.

Starting with an $n=4$ solution, which preserves $1/16$ of the supersymmetry, 
the special 
triple overlaps contained in it 
can be used to construct new solutions with $n>4$ also preserving $1/16$ of the 
supersymmetry.
To illustrate this procedure consider starting with a solution
with four fivebranes,
in which there are three separate 
triple overlaps of dimension two.  This solution can then be extended three 
times to give a configuration with a total of seven fivebranes
and two overall transverse dimensions.  We expect, although it would be
tedious to check,
that this gives a new solution preserving $1/16$ of the
supersymmetry
with fivebranes oriented as follows,
\eqn\seven{\eqalign{&(1,2,3,4,5),\qquad (3,4,5,6,7),\qquad (1,2,3,6,7),\qquad
(1,3,4,6,8),\cr
& (2,3,4,7,8),\qquad (1,3,5,7,8),\qquad (2,3,5,6,8) ,}}
all overlapping in a common 
string.
The polarizations of the last three sets of branes would be determined by 
those of the first four.
As for the other cases with two overall transverse dimensions, this would 
reduce to give black hole type solutions in $D=3$. 
The $2\cap 2\cap 5\cap 5$ given in \klebanov\ 
is the only case of $n\ge 4$ 
overlapping branes with three overall transverse 
directions. Note that a membrane in the $(3,9)$ plane can be added to \seven\ to give a
solution with $n=8$ preserving only $1/32$ of the supersymmetry.

\newsec{Reduction to $D$=4}
The new $D$=11 solutions with fall off in at least three overall transverse
directions
may be dimensionally reduced to give black hole
solutions in $D$=4.  Membranes and fivebranes in $D$=11 give electrically and
magnetically charged black holes in $D$=4, respectively.
Branes aligned along different eleven-dimensional
hyperplanes carry charge under different four-dimensional gauge
fields. Write the $D$=11 metric as
\eqn\reduction{\eqalign{
d\tilde{s}^2 = & e^{-\Phi}g_{\mu\nu}dx^\mu dx^\nu +
\sum_i e^{2\varphi_i}\left(dx^i\right)^2,\cr
\Phi=&\sum_i \varphi_i ,\qquad \mu,\nu=0,\dots,3, \qquad i=4,\dots,10, \cr
}}
with all the fields independent of the $x_i$.
Then $g_{\mu\nu}$ is the $D$=4 Einstein metric, which for the overlapping brane
solutions with $n$=1,2,3 is
given by
\eqn\fourmetric{
ds^2=-\prod_{i=1}^n f_i^{-1/2}dt^2 + \prod_{i=1}^n f_i^{1/2}
\left(dx_1^2+dx_2^2 +dx_3^2\right). }
In the intersecting cases in which the functions $f_i$ are
equal, the metric \fourmetric\ is that of a single dilaton black
hole \dilatonbh\ with dilaton coupling $a=\sqrt{(4/n)-1}$, as noted in \george
{}.
We briefly illustrate how this comes about below, by considering the
dimensionally reduced action in two examples.

Reduce the gauge field by taking only components of the form
$\tilde{F}_{ij\mu\nu}$ to be nonzero. This gives a collection of $21$ two-form
field strengths in $D$=4, which we label $F_{(ij)}$, with $i<j$.
We further restrict to field configurations satisfying
\eqn\vanish{F_{(ij)}\wedge F_{(kl)}=0,}
for all $i,j,k,l$ in order for the cubic term in the $D$=11 action to vanish.
The reduced action is then given by
\eqn\reducedaction{
S=\int\sqrt{-g}\left\{ R-\half\left(\nabla\Phi\right)^2 -
\sum_k\left(\nabla\varphi_k\right)^2 -
\sum_{j<k}F_{(jk)}^2e^{\Phi-2\varphi_j -2\varphi_k}\right\}. }

\subsec{n=1}
For $n$=1, assume that only one of the $21$  $D$=4 field strengths, $F_{(45)}$,
is nonvanishing and satisfies \vanish .
The equations of motion can then be used
to relate all seven scalar fields $\varphi_i$
to a single scalar field $\psi$ according to
\eqn\scalars{
\varphi_4=\varphi_5=\psi,\qquad \varphi_6=\dots =\varphi_{10}=\Phi=-{\psi\over
2}.}
Substituting into \reducedaction\
and rescaling $\psi = {4\over 3\sqrt{3}}\phi$ then gives the standard form of
the
dilaton gravity  action
\eqn\dilatonaction{
S=\int d^4x \sqrt{-g}\left\{ R-2(\nabla\phi)^2 -e^{-2a\phi}F^2\right\}, }
with $a=\sqrt{3}$.

\subsec{n=2}
Consider now the
$n$=2 membrane (fivebrane) solutions in which only two of the four dimensional
field strengths,  $F_{(45)},F_{(67)}$, are nonzero.  Since these are pure
electric (magnetic),  the condition \vanish\ is satisfied.
The equations of motion allow a truncation to two scalar fields
$\psi$ and $\eta$ given by
\eqn\doublescalars{
\varphi_4=\varphi_5=\psi,\qquad \varphi_6=\varphi_7=\eta,\qquad
\varphi_8=\varphi_9=\varphi_{10}=
\Phi = -\psi-\eta. }
Further taking the combinations $\lambda=\psi +\eta$ and $\Omega=\psi-\eta$,
the reduced action
\reducedaction\ becomes
\eqn\doubleaction{S=\int\sqrt{-g}\left\{ R- {9\over 2}(\nabla\psi)^2 -
(\nabla\Omega)^2
- e^{-(3\lambda+2\Omega)}
F_{(45)}^2 -e^{-(3\lambda-2\Omega)}F_{(67)}^2\right\}
.}
In the intersecting case with $f_1=f_2$,
$F_{45}=F_{67}$ and we see that the source for
$\Omega$
vanishes.  Setting $\Omega =0$ and rescaling  $\lambda={2\over 3}\phi$
then gives the
standard dilaton gravity action with $a=1$.  Similar considerations show how
the $a=\sqrt{(4/n)-1}$ dilaton gravity action arises in the  remaining $n$=2,3
intersecting cases in which the functions $f_i$ are taken to be equal.

\newsec{More Overlapping Branes}
The overlapping brane solutions we have considered so far
are all translationally invariant in the relative transverse
directions.
As we pointed out earlier, this suggests that these
solutions be considered after dimensional reduction in the
relative transverse directions. It is natural to ask if
``true" overlapping brane solutions exist where this restriction is
relaxed.
In ten dimensions
the answer turns out to be yes.
Specifically, one can have two 5-branes
that overlap in a string. In fact this type of solution was first constructed
by Khuri in \khuri\ (see also \kounnas)
but the interpretation as overlapping branes
was not discussed. After first considering new solutions generated
from this solution by $T$-duality we will see that there exists a new
solution of $M$-theory describing two fivebranes overlapping in a string.

By setting the gauge fields to zero in the heterotic solution in \khuri\
one gets type IIA and type IIB solutions
describing two NS-NS 5-branes overlapping in a string:
\eqn\nsnsffs{
\eqalign{
ds^2=&-dt^2+dx_1^2+
f_2(dx_2^2+dx_3^2+dx_4^2+dx_5^2)
+f_1(dx_6^2+dx_7^2+dx_8^2+dx_9^2)\cr
H_{mnp}=&-{c_1\over 2}\epsilon_{mnpq}\p_qf_1
\qquad H_{\mu\nu\lambda}=-{c_2\over 2}\epsilon_{\mu\nu\lambda\rho}
\p_\rho f_2,\qquad e^{2\phi}=f_1 f_2\cr
f_1=&f_1(x_6,x_7,x_8,x_9), \qquad f_2=f_2(x_2,x_3,x_4,x_5),\qquad
\nabla^2f_i=0,
\cr}
}
where $x^\mu, \mu=2,\dots 5$ are the spatial coordinates of one 5-brane,
$x^m, m=6,\dots 9$ are the spatial coordinates on the second brane and
the epsilon tensors are those of the corresponding flat space.
Note that the harmonic function of each brane has non-trivial falloff
in the spatial directions of the other brane. As shown in \khuri\ the
solution preserves $1/4$ of the supersymmetry.

Using $SL(2,Z)$ duality of type IIB theory we can generate a solution
describing two $D$-5-branes overlapping in a string:
\eqn\nsnsffs{
\eqalign{
d\bar s^2=&(f_1f_2)^{-1/2}(-dt^2+dx_1^2)+
f_1^{-1/2}f_2^{1/2}(dx_2^2+dx_3^2+dx_4^2+dx_5^2)\cr
&+f_1^{1/2}f_2^{-1/2}(dx_6^2+dx_7^2+dx_8^2+dx_9^2),
\qquad e^{2\bar \phi}=f_1^{-1} f_2^{-1}\cr
H'_{mnp}=&-{c_1\over 2}\epsilon_{mnpq}\p_qf_1
\qquad H'_{\mu\nu\lambda}=-{c_2\over 2}\epsilon_{\mu\nu\lambda\rho}
\p_\rho f_2\cr
f_1=&f_1(x_6,x_7,x_8,x_9), \qquad f_2=f_2(x_2,x_3,x_4,x_5),\qquad
\nabla^2f_i=0,
\cr}
}
where the $H'$ is the R-R three form field strength.
Acting now with $T$-duality we can generate a set of overlapping brane
solutions of type II supergravity given by
\eqn\listt{
\eqalign{IIA:\qquad &0\cap 8 (0);\quad 2\cap 6 (0);\quad 2\cap 8 (1);
\quad 4\cap 4 (0);
\quad 4\cap 6 (1);\cr
IIB:\qquad &1\cap 7 (0);\quad 1\cap 9 (1);
\quad  3\cap 5 (0);\quad  3\cap 7 (1);\quad  5\cap 5 (1);\cr
}
}
It is worth pointing out that this list corresponds
at a string theory level to superpositions of branes with eight $ND$ directions
and thus, according to \dbranenotes, should be supersymmetric. It is not
hard to convince oneself that \listt\ provides a complete list of cases
involving two branes with
eight $ND$ directions.

To obtain the new $D$=11 solution describing two fivebranes overlapping
in a string, we can uplift
the $4\cap4 (0)$ solution of type IIA. Alternatively we can uplift
the NS-NS overlapping 5-branes \nsnsffs\ directly.
In either case we are led to the
solution
\eqn\mtffs{
\eqalign{
d\tilde s^2=&(f_1f_2)^{-1/3}(-dt^2+dx_1^2)+
f_1^{-1/3}f_2^{2/3}(dx_2^2+dx_3^2+dx_4^2+dx_5^2)\cr
&+f_1^{2/3}f_2^{-1/3}(dx_6^2+dx_7^2+dx_8^2+dx_9^2)
+f_1^{2/3}f_2^{2/3}dx_{10}^2\cr
\tilde F_{mnp10}=&-{c_1\over 2}\epsilon_{mnpq}\p_qf_1,\qquad
\tilde F_{\mu\nu\lambda10}= -{c_2\over 2}\epsilon_{\mu\nu\lambda\rho}
\p_\rho f_2.
\cr
}
}
The solution preserves $1/4$ of the $D$=11 supersymmetry. The Killing spinors
are of the form $\epsilon=(f_1f_2)^{-1/12}\eta$ with the constant spinor
$\eta$ satisfying
\eqn\again{
\eqalign{
\hat\Gamma^6\hat\Gamma^7\hat\Gamma^8\hat\Gamma^9\hat\Gamma^{10}=&c_1\epsilon
\cr
\hat\Gamma^2\hat\Gamma^3\hat\Gamma^4\hat\Gamma^5\hat\Gamma^{10}=&c_2\epsilon.
\cr
}
}
This solution describes two fivebranes in the $(1,2,3,4,5)$
and $(1,6,7,8,9)$ directions overlapping
in a string in the $(1)$ direction.
Note that although the functions do depend on the relative transverse
directions, they are translationally invariant in the
overall transverse direction $x_{10}$. So
unlike the $D$=10
solution \nsnsffs\ this overlapping solution is not a true overlap.

\newsec{Conclusion}
In this paper we have constructed a class of supersymmetric
solutions of $D$=11 supergravity describing $n$=2,\dots ,8 
overlapping membranes and
fivebranes.  The pairwise overlaps of the branes in this class can either be:
two fivebranes in a threebrane, a fivebrane and a membrane in a string or
two membranes in a point.
The solutions preserve at least $2^{-n}$ 
of the supersymmetry and generalize those
presented in \george,\guven.  For the cases $n=2,3$ the solutions preserve
exactly $2^{-n}$ of the supersymmetry.
We explicitly showed that the $n$=2 solutions and certain of the $n$=3 
solutions reduce to collections of extremal dilaton black holes in $D$=4.

In section (5) we constructed an additional
$D$=11 solution describing two
fivebranes intersecting in a string that preserves $1/4$ of the supersymmetry.
This solution includes non-trivial dependence on the relative transverse
coordinates. It would be interesting to know if this feature can be 
further generalised in this case and 
to other configurations of membranes and fivebranes.

By reducing the $D$=11 solutions to $D$=10 and using $T$-duality we also showed
how a large class of solutions of type IIA and IIB supergravity could
be constructed.  It would be interesting to further study these 
and additional solutions involving NS branes, which may be obtained by acting
with $SL(2,Z)$. Finally, one may also try to generalize the overlapping
solutions to include travelling waves along the branes as in
\garf,\dghw,\callan.

\bigskip
\noindent{\bf Acknowledgements}

We thank  F. Dowker for useful
discussions and K.Z. Win for providing the simple formula in \win\ for
computing the curvature of diagonal metrics.  JPG is supported in part by
the U.S. Dept. of Energy under Grant No. DE-FG03-92-ER40701.  JT is supported
in
part by NSF grant  NSF-THY-8714-684-A01.

\listrefs
\end